\documentclass[a4paper,12pt]{article}
\usepackage{amsmath,amsfonts,amssymb}
\usepackage{cite}

\newcommand{\be}{\begin{equation}}
\newcommand{\ee}{\end{equation}}
\newcommand{\ba}{\begin{eqnarray}}
\newcommand{\ea}{\end{eqnarray}}
\newcommand{\ci}[1]{\cite{#1}}
\newcommand{\nl}{\nonumber \\}
\newcommand{\Lgg}[1]{\ln^{#1}\!\frac{\Lambda^2}{M_0^2}}
\def\gl#1{(\ref{#1})}
\newcommand{\ac}[1]{O\left(
\frac{1}{\ln^{#1}\!\frac{\Lambda^2}{M_0^2}}\right)}

\begin{document}

\begin{center}
{\Large\bf Vector particles in Quasilocal Quark
Models}\footnote{Published in Zap. Nauch. Sem. POMI {\bf 291}
(2002), 5} \\

\vspace{4mm}

V. A. Andrianov and S. S. Afonin \\
V. A. Fock Department of Theoretical Physics, St. Petersburg State
University,
198504 St. Petersburg, Russia
\end{center}

\renewcommand{\abstractname}{Abstract}

\begin{abstract}
We consider the Quasilocal Quark Model of NJL type including vector and
axial-vector four-fermion interaction with derivatives. The mass
spectrum for the ground and first excited states is obtained.
The chiral symmetry restoration sum rules in these
channels are imposed as matching rules to QCD at intermediate energies and a
set of constraints on parameters of QQM is performed.
\end{abstract}

\section{Introduction}


The main purpose of this paper is to describe the physics of as
ground vector (V) and axial-vector (A) meson resonances at
low energies as excited states with increasing masses at
intermediate energies in the framework of effective action of
Quasilocal Quark Models (QQM) which were introduced in~\ci{11}.
We will see that these effective QCD QQM are sufficiently general
and from the physical point of view evident to allow a good
description of wide set of mass relations for vector and
axial-vector mesons. Moreover, using properties of two-point QCD
VA correlators of quark densities at high energies (asymptotic freedom
and Operator Product Expansion (OPE)) and two-resonance ansatz for
VA meson correlators of QQM, we will obtain a number of
constraints for parameters of QQM from the so-called Chiral
Symmetry Restoration (CSR) sum rules.

It is well known that the effective QCD-inspired quark models with
four-fermion interactions of Nambu-Jona-Lasinio (NJL) type are often
applied to describe the
low-energy QCD in the hadronization regime~\ci{1,2,3,4,5,6,7,8,9,10}.
The appearance of
effective quark theory is connected with one of specific
property of QCD --- the dynamical Chiral Symmetry Breaking (CSB).
The local four-fermion interaction is involved to induce the
CSB due to strong attraction in the
scalar channel. As a consequence, the dynamical quark mass
$m_{\mbox{\scriptsize dyn}}$ is created, as well as an isospin multiplet of pions,
massless in the chiral limit, and a massive scalar meson with mass
$m_{\sigma}=2m_{\mbox{\scriptsize dyn}}$ appear. However, it is known from
experiment that there are series of meson states
with equal quantum numbers~\cite{pdg}. 
Due to confinement, one expects an infinite number of such excited
states with increasing masses.
In such a way, the problem arises how to describe the physics of
those resonances at intermediate energies. To solve this problem one can
extend the NJL quark model
taking into account  higher-dimensional quark operators
with derivatives, i.e. quasilocal quark interactions~\ci{11}. For
sufficiently strong couplings the new
operators promote the formation of additional meson states.
Such a
quasilocal approach (see also~\ci{12,13,14,volw,volkeb}) represents a
systematic extension of the NJL-model towards the complete
effective action of QCD where many-fermion vertices with
derivatives possess the manifest chiral symmetry of
interaction, motivated by the soft momentum expansion of the
perturbative QCD effective action.
In the effective action of Quasilocal Quark Models (QQM) of NJL type
the low-energy
gluon effects are  hidden in the coupling constants.
The alternative schemes including the condensates of low-energy
gluons can be found in~\ci{6,15}.

At the same time in the large-$N_c$ approach, which is equivalent
to planar QCD~\cite{16},
the correlators of color-singlet quark currents are saturated
by narrow meson resonances. In particular, the two-point
correlators of 
vector and axial-vector
quark currents are represented by the sum of related meson poles
in Euclidean space:
\be
\Pi^{C}(p^{2})=\int d^{4}x\exp(ipx)\langle \bar{q}\Gamma
q(x)\bar{q}\Gamma q(0)
\rangle=\sum_{n}\frac{Z_{n}^{C}}{p^{2}+m_{C,n}^{2}}
+D_{0}^{C}+D_{1}^{C}p^{2}, \label{cor1}
\ee
$$
C\equiv 
V,A; \qquad
\gamma_{\mu},\gamma_{\mu}\gamma_{5}; \qquad
D_{0},D_{1}=const.
$$
The last two terms 
represent a perturbative contribution,
with $D_{0}$ and $D_{1}$ being contact terms required for the
regularization of infinite sums. On the other hand,
their high-energy asymptotics is
provided~\cite{18} by the perturbation theory and
OPE due to asymptotic freedom of QCD. Therefrom the
above correlators increase at large $p^{2}$,
\be
\Pi^{C}(p^{2})\mid_{p^{2}\rightarrow\infty}\sim
p^{2} \ln\frac{p^{2}}{\mu^{2}}\,.
\ee
When comparing the two expressions above one concludes that the
infinite series of resonances with the same quantum numbers
should exist in order to reproduce the perturbative asymptotics.
So at intermediate energies the correlators of QQM can be matched~\cite{zap}
to the OPE of QCD correlators. This matching realizes the
correspondence to QCD and improves the predictability of QQM.

Meantime the differences of correlators of opposite-parity currents
rapidly decrease at large momenta $p^{2}\rightarrow\infty$ 
\cite{18,20,21}
\be
\Pi^{V}(p^{2})-\Pi^{A}(p^{2})\equiv
\frac{\Delta_{V\!\!A}}{p^{6}}-\frac{m_0^2\Delta_{V\!\!A}}{p^{8}}+
O\left(\frac{1}{p^{10}}\right),\,\, \Delta_{V\!\!A}\simeq
-16\pi\alpha_{s}\langle\bar{q}q\rangle^{2}, \label{VA}
\ee
where $m_0^2=0.8\pm0.2\,\mbox{GeV}^2$~\ci{iof}, $\alpha_s\approx0.3$
(at 1 GeV), and we have defined in the $V,A$ channels
\be
\Pi_{\mu\nu}^{V,A}(p^{2})\equiv(-\delta_{\mu\nu}p^{2}+p_{\mu}p_{\nu})
\Pi^{V,A}(p^{2}).
\ee

Therefore the chiral symmetry is restored at high energies and
the
difference~\gl{VA} represents an order parameter of CSB in QCD. As it
decreases rapidly at large momenta
one can perform the matching of QCD asymptotics
by means of few lowest lying resonances that gives a set of
constraints for QQM parameters from the CSR sum rules.

In the present work the vector QQM model is considered with two channels
where two pairs of vector and axial-vector mesons are
generated. Respectively it is expected to reproduce the lower
part of QCD meson spectrum in the planar limit and the leading asymptotics
of chiral symmetry restoration for higher energies.
In the Sec.~2 we define the VA, $SU(2)$ QQM with two pairs of
VA-mesons and the corresponding mass spectrum for boson states is obtained.
The Sec.~3 is devoted to the $U(3)$ extension of VA QQM.
With the help of two-resonance ansatz for VA-mesons the correlators of QQM
are matched to the OPE of QCD correlators and a number of
constraints for parameters of QQM from CSR sum rules are performed
in the Sec.~4.

\section{VA, $SU(2)$ Quasilocal Quark Model}

The $SU(2)$ QQM Lagrangian for the two-channel vector (V) and axial-vector
(A) case in the chiral limit~$m_q = 0$ has the form~\ci{we1} (in
Euclidean space):
\be
\label{su2}
{\cal L}_{V\!\!A}=\bar{q}i\!\!\!\not{\partial\phantom{d}}\!\!\!q+
\frac{1}{4N_{f}N_{c}\Lambda^{2}}\sum_{k,l=1}^{2}
b_{kl}\left[\bar{q}\Gamma_{V,k}^iq\cdot\bar{q}\Gamma_{V,l}^iq+
\bar{q}\Gamma_{A,k}^iq\cdot\bar{q}\Gamma_{A,l}^iq\right],
\ee
$$
\Gamma_{V,k}^i\equiv i\gamma_{\mu}f_k(s)\tau^i,\qquad
\Gamma_{A,k}^i\equiv i\gamma_{\mu}\gamma_5f_k(s)\tau^i,\qquad
i=1,2,3\,,
$$
where $q\equiv q_j$ ($j$ is the number of flavor $N_f$) are color
fermionic fields with $N_c$ components, $b_{kl}$ represents the
symmetric nonsingular matrix of real coupling constants, and
$\tau^i$ denote Pauli matrices. The quantities $f_k( s)$,
$s\rightarrow -\partial^2/\Lambda^2$ are the form factors
specifying the quasilocal interaction. We accept the following
sequence of action of the derivatives for the Hermitian fermion
currents:
\be
\bar q\frac{\partial^2}{\Lambda^2} q=\frac14\bar q
\biggl(\frac{\stackrel{\rightarrow}{\partial}-
\stackrel{\leftarrow}{\partial}}{\Lambda}\biggr)^2 q\, .
\ee
In addition, let us regularize the interaction vertices by
introducing the momentum cutoff
\be
\bar q q\longrightarrow\bar q\theta(\Lambda^2+\partial^2) q\,,
\label{5}
\ee
and choose the polynomial form factors  as being orthogonal on
the unit interval,
\be
\int_{0}^{1}f_{k}(s)f_{l}(s)ds=\delta_{kl}\,.
\ee
We select out here:
\be
f_{1}(s)=2-3s\,, \qquad f_{2}(s)=-\sqrt{3}s\,.
\ee
As this model interpolates the low-energy QCD action
it is supplied with a cutoff $\Lambda$ (of order of the
CSB scale) for virtual quark momenta in quark loops.

It is convenient to pass to the auxiliary vector ($\rho_{\mu}^i$) and
axial-vector ($a_{\mu}^i$) fields,
\ba
{\cal L}_{\mbox{\scriptsize aux}}&=&\bar{q}i\!\!\!\not{\partial\phantom{d}}\!\!\!q+
\sum_{k=1}^{2}i\bar{q}\left(\Gamma_{V,k}^i\rho_{k,\mu}^i+
\Gamma_{A,k}^i a_{k,\mu}^i\right)q+ \nl
&+&N_fN_c\Lambda^{2}\sum_{k,l=1}^{2}\left(\rho_{k,\mu}^ib_{kl}^{-1}
\rho_{l,\mu}^i+a_{k,\mu}^ib_{kl}^{-1}a_{l,\mu}^i\right).
\ea

After integrating out the quark fields
$$
\biggl\langle \exp\biggl(-\int\! d^{4}x\,{\cal L}\biggr)\biggr
\rangle_{\bar q q}\equiv\exp(- S_{\mbox{\scriptsize eff}}),
$$
one comes to the bosonic effective action:
$$
S_{\mbox{\scriptsize eff}}(\rho_{\mu,k}^i, a_{\mu,k}^i)=N_fN_c\Lambda^2
\sum_{k,l=1}^{2}\{\rho_{k,\mu}^ib_{kl}^{-1}\rho_{l,\mu}^i+
a_{k,\mu}^ib_{kl}^{-1}a_{l,\mu}^i\}-N_fN_c\mbox{Tr}\ln
\left.\lefteqn{\not{\phantom{DD}}}D\right|_{\mbox{\scriptsize reg}}\,,
$$
\be
\label{eff2}
\lefteqn{\not{\phantom{DD}}}D\equiv i(\!\not{\partial}+M)+i\sum_{k=1}^{2}
\left(\Gamma_{V,k}^i\rho_{k,\mu}^i+\Gamma_{A,k}^i a_{k,\mu}^i\right),
\ee
where we have introduced the dynamic mass function
$M\equiv\sum_k\sigma_kf_k(s)$, with $\sigma_k$ being the vacuum
expectation values of scalar fields~\ci{24}.
We use the chirally invariant regularization of fermionic
determinant:
$$
\ln\det\lefteqn{\not{\phantom{DD}}}D=\mbox{Tr}^{\mbox{\scriptsize all}}
\ln\lefteqn{\not{\phantom{DD}}}D
\longrightarrow N_fN_c\mbox{Tr}\ln\left.\lefteqn{\not{\phantom{DD}}}D
\right|_{\mbox{\scriptsize reg}}\equiv\frac12N_fN_c\mbox{Tr}\ln\frac{\lefteqn{\not{\phantom{DD}}}D
\lefteqn{\not{\phantom{DD}}}D^{\dag}}{\mu^2}\,,
$$
where the constant $\mu$ is a normalization scale for quark
fields and the trace "Tr" is assumed over all degrees of freedom except
color and flavor ones. We will carry out further analysis in the mean
field approximation ($N_c\gg~1$).
Expanding the Eq.~\gl{eff2} in boson fields and retaining the quadratic
in fields part only, one obtains:
\be
S_{\mbox{\scriptsize eff}}^{(2)}(\rho_{\mu,k}^i, a_{\mu,k}^i)=\frac{1}{2}\int\frac{d^{4}p}
{(2\pi)^{4}}\sum_{k,l=1}^{2}
[\rho_{k,\mu}^iC_{kl}^{\rho,\mu\nu}\rho_{l,\nu}^i+a_{k,\mu}^i
C_{kl}^{a,\mu\nu}a_{l,\nu}^i].
\ee
The inverse propagators are defined by the corresponding second
variation of $S_{\mbox{\scriptsize eff}}(\rho_{\mu,k}^i, a_{\mu,k}^i)$:
$$
C_{kl}^{(\rho,a)\mu\nu}=2N_fN_c\Lambda^2b_{kl}^{-1}\delta_{\mu\nu}-
N_fN_c\int\frac{d^4q}{(2\pi)^4}\times
$$
\be
\label{secvar}
\mbox{tr}\left\{(i\gamma_{\mu},i\gamma_{\mu}\gamma_{5})\frac{f_{k}
\left(\left(\frac{q+p/2}{\Lambda}\right)^2\right)}{\not
q+\frac{1}{2}\!\!\not p+iM}
(i\gamma_{\nu},i\gamma_{\nu}\gamma_{5})\frac{f_{l}
\left(\left(\frac{q-p/2}{\Lambda}\right)^2\right)}{\not
q-\frac{1}{2}\!\!\not p+iM}\right\},
\ee
where the trace "tr" spans the Dirac indices only.

To obtain mass spectrum we expand~(\ref{secvar}) in a small external
momentum $p$ ($p^2/\Lambda^2\ll 1$) up to terms $\sim\! p^2$ and
calculate the corresponding loop integral using the momentum
cutoff regularization. To compensate the quadratic divergences in
this integral we parametrize the matrix of coupling constants as
follows:
\be
\label{ft}
16\pi^{2}b_{kl}^{-1}=\delta_{kl}-\frac{4}{3}
\frac{\bar\Delta_{kl}}{\Lambda^{2}}\,;
\qquad \bar\Delta_{kl}\ll\Lambda^{2}.
\ee
The general structure of~\gl{secvar} is:
\be
\label{varder}
C_{kl}^{(\rho,a)\mu\nu}=
\frac{N_fN_c}{12\pi^2}\left[\left(\hat{A}_{kl}^{(\rho,a)}p^2+
\hat{B}_{kl}^{(\rho,a)}\right)
\delta_{\mu\nu}-\hat{A}_{kl}^{(\rho,a)}p_{\mu}p_{\nu}\right]
+O\left(\frac{1}{\Lambda^2}\right),
\ee
where the two symmetric matrices - the kinetic term $\hat{A}$ and
the momentum independent part $\hat{B}$ - have been introduced:
\be
\label{aa}
\hat{A}^{(\rho,a)}\equiv\begin{pmatrix}
  4\ln\!\frac{\Lambda^{2}}{M_{0}^{2}}-\frac{15}{2} & -\frac{\sqrt3}{2} \\
  -\frac{\sqrt3}{2} & \frac{3}{2}
\end{pmatrix},
\ee
\be
\label{br}
\hat{B}^{\rho}\equiv\begin{pmatrix}
  -2\bar{\Delta}_{11} & -2\bar{\Delta}_{12} \\
  -2\bar{\Delta}_{12} & -2\bar{\Delta}_{22}
\end{pmatrix},
\ee
\be
\label{ba}
\hat{B}^{a}\equiv\begin{pmatrix}
  -2\bar{\Delta}_{11}+\sigma_{11}
   & -2\bar{\Delta}_{12}+\sigma_{12}
    \\
  -2\bar{\Delta}_{12}+\sigma_{12}
   & -2\bar{\Delta}_{22}+\sigma_{22}
\end{pmatrix},
\ee
$$
\sigma_{11}\equiv 24M_{0}^{2}\ln\!\frac{\Lambda^{2}}{M_{0}^{2}}-
\frac{477}{2}\sigma_{1}^{2}
-15\sqrt{3}\sigma_{1}\sigma_{2}+\frac{9}{2}\sigma_{2}^{2}\,,
$$
$$
\sigma_{12}\equiv -\frac{15\sqrt{3}}{2}\sigma_{1}^{2}+9\sigma_{1}\sigma_{2}
+\frac{3\sqrt{3}}{2}\sigma_{2}^{2}\,,
$$
$$
\sigma_{22}\equiv  \frac{9}{2}\sigma_{1}^{2}+3\sqrt{3}\sigma_{1}\sigma_{2}+
\frac{27}{2}\sigma_{2}^{2}\,.
$$
Here $M_0\equiv M(0)=2\sigma_1$ is the dynamic quark mass at zero
external momentum. The remaining logarithmic divergences will be
absorbed later by meson masses and renormalization of meson
fields.

The physical mass spectrum is defined by the secular equation
\be
\label{sec}
\det(\hat{A}p^{2}+\hat{B})^{(\rho,a)}=0\,,\qquad
m_{phys}^2=-p_0^2\,.
\ee
As it will be seen further the consistency with CSR sum rules
imposes the following scale condition:
\be
\label{scale}
\bar\Delta_{kl}=O\left(\Lambda^2\right).
\ee
Using~\gl{aa},~\gl{br} one has for the Eq.~\gl{sec}:
$$
6\left(\ln{\!\frac{\Lambda^{2}}{M_{0}^{2}}}-2\right)p^4-
\left(8\ln{\!\frac{\Lambda^{2}}
{M_{0}^{2}}}\,\bar{\Delta}_{22}-15\bar{\Delta}_{22}+3\bar{\Delta}_{11}+
2\sqrt{3}\bar{\Delta}_{12}\right)p^2+
$$
\be
\label{sec2}
+4\,\det\hat{\bar{\Delta}}=0,
\ee
The solution of Eq.~\gl{sec2} in the large-log approximation
$\ln\!\frac{\Lambda^{2}}{M_{0}^{2}}\gg 1$ 
is as follows:
\be
\label{mr1}
m_{\rho}^2=-
\frac{\det\hat{\bar{\Delta}}}{2\ln{\!\frac{\Lambda^{2}}{M_{0}^{2}}}\,
\bar{\Delta}_{22}}+c_1+\ac{} ,
\ee
\be
\label{mr2}
m_{\rho'}^2=-\frac43\bar{\Delta}_{22}+\delta+c_2+\ac{} .
\ee
To obtain the A-meson mass spectrum it is enough to make the
replacement (see \gl{br}, \gl{ba}):
$\bar{\Delta}_{kl}\rightarrow\bar{\Delta}_{kl}-1/2\sigma_{kl}$.
The result is:
\be
\label{-ma1}
m_{a_{1}}^2=-\frac{\det\hat{\bar{\Delta}}}{2\ln{\!\frac{\Lambda^{2}}
{M_{0}^{2}}}\,\bar{\Delta}_{22}}+
6M_0^2+c_1+\ac{},
\ee
\be
\label{-ma2}
m_{a_{1}'}^{2}=
-\frac43\bar{\Delta}_{22}+3\bar{\sigma}+\delta
+c_2+\ac{}.
\ee
The prime labels everywhere the corresponding excited meson state
and we have introduced the notations:
\be
\label{del}
\delta\equiv-\frac{6m_{\rho}^2\Lgg{}+d}{6\Lgg{}}\,,
\quad c_1 \sim c_2 = O\left(\frac{\Lambda^2}{\Lgg{2}}\right),
\ee
\be
\label{d}
d\equiv3\bar{\Delta}_{11}+2\sqrt{3}\bar{\Delta}_{12}+\bar{\Delta}_{22}\,,
\quad \bar{\sigma}\equiv\sigma_{1}^{2}+
\frac{2\sqrt{3}}{3}\sigma_{1}\sigma_{2}+3\sigma_{2}^{2}>0\,.
\ee
As it is seen from the Eqs.~\gl{mr1}-\gl{-ma2} the scale of mass
squared for ground VA states is $O(\Lambda^2/\Lgg{})$ and for
excited ones is $O(\Lambda^2)$. Thus, the excited states turn out to be
logarithmically heavier than ground ones as it was for the scalar (S) and
pseudoscalar (P) case~\ci{24}.
This qualitative property is independent of any concrete choice of
form factors.
Combining the Eqs.~\gl{mr1}-\gl{-ma2} with the corresponding
results 
in~\ci{24} one obtains:
\be
\label{ma1}
m_{a_1}^{2} - m_{\rho}^{2} = 6M_{0}^{2}+\ac{}=\frac32m_{\sigma}^2+\ac{},
\ee
\be
\label{ma2}
m_{a_1'}^{2} - m_{\rho'}^{2} =
3\bar{\sigma}+\ac{}=\frac32(m_{\sigma'}^2-
m_{\pi'}^2)+\ac{}.
\ee
The last equalities in Eqs.~(\ref{ma1}),~(\ref{ma2}) do not depend
on model parameters. Note also, that differences of masses
squared both in the Eq.~(\ref{ma1}) and in the Eq.~\gl{ma2} are of order
$O(1)$, which indicates the chiral symmetry restoration at a scale
over 1~GeV.

Let us comment the approximations used to
derive the meson mass spectrum: namely, the large $N_c$
and leading-log approximations. The first one is equivalent
\cite{16} to the neglect of
meson loops. The second one fits well the quarks
confinement  as
quark-antiquark threshold contributions are suppressed in two-point
functions in the leading
log approximation. The accuracy of
this approximation is controlled also  by the magnitudes of higher dimensional
operators neglected in QQM, i.e. by contributions of heavy
mass resonances not included into QQM. All these approximations
are mutually consistent. In particular, in the effective action without
gluons the quark confinement should be realized with the help of
an infinite number of quasilocal vertices with higher-order derivatives.
Then the imaginary part of quark loops can be compensated and
their momentum dependence can eventually reproduce the infinite sum of meson
resonances in
the large-$N_c$ limit. If the effective action is truncated with a finite
number of vertices and thereby deals
with only a few resonances  one has to retain only a finite number of
terms in the low-momentum expansion of quark loops in the CSB phase,
with a non-zero dynamic mass.

\section{VA, $U(3)$ extension of Quasilocal Quark Model}

In this section we build $U(3)$ version of QQM~\ci{we3}, with current quark
masses being taken into account. The corresponding extension of the
VA, $SU(2)$ QQM Lagrangian~\gl{su2} is:
\be
\label{su3}
{\cal
L}_{V\!\!A}\!=\!\bar{q}i(\hat{\not{\partial\phantom{d}}}\!\!\!+\hat{m})q\!+\!
\frac{1}{4N_{f}N_{c}\Lambda^{2}}\!\sum_{k,l=1}^{2}\!
b_{kl}^i\left[\bar{q}\Gamma_{V,k}^iq\cdot\bar{q}\Gamma_{V,l}^iq\!+\!
\bar{q}\Gamma_{A,k}^iq\cdot\bar{q}\Gamma_{A,l}^iq\right]\!,
\ee
$$
\Gamma_{V,k}^i\equiv i\gamma_{\mu}f_k(s)\lambda^i,\qquad
\Gamma_{A,k}^i\equiv i\gamma_{\mu}\gamma_5f_k(s)\lambda^i,\qquad
i=0,...,8\,,
$$
where $\lambda^i$ represent Gell-Mann matrices. The current quark mass
matrix is $\hat{m}=diag(m_u,m_d,m_s)$.
In the sequel we adopt the exact isospin symmetry $m_u=m_d$.
The symbol $\tilde{u}$ will stand everywhere for the $u$, $d$, $\bar{u}$,
$\bar{d}$ quarks. The symbol $\tilde{s}$ will denote $s$ or $\bar{s}$
quarks. The generalization of fine-tuning condition~\gl{ft} takes
the form:
\be
16\pi^{2}(b_{kl}^i)^{-1}=\delta_{kl}^i-\frac{4}{3}
\frac{\bar\Delta_{kl}^i}{\Lambda^{2}}\,;
\qquad \bar\Delta_{kl}^i\ll\Lambda^{2}.
\ee
The couplings $\bar{\Delta}_{kl}^i$ satisfy the relations
\be
\label{par2}
\bar{\Delta}_{kl}^m=\bar{\Delta}_{kl}^0\,,\quad\bar{\Delta}_{kl}^n=
\frac12\left(\bar{\Delta}_{kl}^0+\bar{\Delta}_{kl}^8\right);\quad
m=1,2,3\,,\quad n=4,5,6,7\,,
\ee
which provide the $U(3)$ Gell-Mann-Okubo relations
\be
\label{gmo}
m_{\alpha,\tilde{u}\tilde{u}}^2+m_{\alpha,\tilde{s}\tilde{s}}^2=
2m_{\alpha,\tilde{s}\tilde{u}}^2\,;\qquad
m_{\alpha,\tilde{u}\tilde{u}}=m_{\alpha,singlet}\,.
\ee
Here $\alpha\equiv V,A,V',A'$. The scheme of calculation of the
mass-spectrum for VA, $U(3)$ QQM is the same as in the Sec.~2 and
the details can be found in~\cite{we3}. We point out only some
features. First, the self-consistency of mass-spectrum imposes the
self-consistency condition~\gl{scale} for all $i$. Second, in the
effective action we did not take into account so far both the
$P-A$ mixing and different mixing terms, which are proportional to
the current quark masses, since these contributions do not
influence on the meson mass spectrum.

The mass relations, which are independent of model parameters in
the large-log approach are:
\be
\label{u28}
m_{a_1,\tilde{u}\tilde{u}}^2-m_{\rho}^2\simeq\frac32\left(
m_{\sigma,\tilde{u}\tilde{u}}^2-m_{\pi}^2\right),\quad
m_{a_1,\tilde{s}\tilde{u}}^2-m_{K^*}^2\simeq\frac32\left(
m_{\sigma,\tilde{s}\tilde{u}}^2-m_{K}^2\right),
\ee
\be
\label{u29}
m_{a_1,\tilde{s}\tilde{s}}^2-m_{\varphi}^2\simeq\frac32\left
[m_{\sigma,\tilde{s}\tilde{s}}^2-\left(
2m_{K}^2-m_{\pi}^2\right)\right].
\ee
\be
m_{a_1',\tilde{u}\tilde{u}}^2-m_{\rho'}^2\simeq\frac32\left(
m_{\sigma',\tilde{u}\tilde{u}}^2-m_{\pi'}^2\right),\quad
m_{a_1',\tilde{s}\tilde{u}}^2-m_{{K^*}'}^2\simeq\frac32\left(
m_{\sigma',\tilde{s}\tilde{u}}^2-m_{K'}^2\right),
\ee
\be
\label{u31}
m_{a_1',\tilde{s}\tilde{s}}^2-m_{\varphi'}^2\simeq\frac32\left(
m_{\sigma',\tilde{s}\tilde{s}}^2-m_{(\eta)'}^2\right).
\ee
The $(\eta)'$ in the Eq.~\gl{u31} is the radial excitation of $\eta$ meson
and throughout the
paper the sign $\simeq$ denotes the large-log approximation.
The corresponding fits and comparisons with experiment are carried
out in~\ci{we3,we2}. The agreement with experimental data is within 10\% for
V-case and 15\% for the A-one. In such a way, the VA, $U(3)$ extension of
QQM provides the certain
 relations between meson masses of multiplets
which are independent of model parameters.

The axial anomaly was not yet considered in framework of QQM and,
thus, we did not include the $\eta'$ meson. Since the ground
P-meson masses are subject to rather $SU(3)$ Gell-Mann-Okubo
relation one has no $\eta$ meson mass in the Eq.~\gl{u29}. On the
other hand, we do not expect a strong $U_A(1)$ anomaly effect for
the case of excited $\eta'$ meson. As a result, the
relations~\gl{gmo} hold for excited P-multiplet, which is
reflected in the Eq.~\gl{u31}.

\section{Chiral symmetry restoration sum rules and constraints on parameters
of VA, $SU(2)$ QQM}

In this section we exploit the constraints based on chiral symmetry
restoration in QCD at high energies for the $SU(2)$ QQM. Expanding
the meson correlators
\gl{cor1} in powers of $p^{2}$ one arrives to the CSR sum rules:
\be
\sum_{n}Z_{n}^{V}-\sum_{n}Z_{n}^{A}=4F_{\pi}^{2}\,,
\ee
\be
\sum_{n}Z_{n}^{V}m_{V,n}^{2}-\sum_{n}Z_{n}^{A}m_{A,n}^{2}=0\,,
\ee
\be
\sum_{n}Z_{n}^{V}m_{V,n}^{4}-\sum_{n}Z_{n}^{A}m_{A,n}^{4}=\Delta_{V\!\!A}\,,
\ee
\be
\sum_{n}Z_{n}^{V}m_{V,n}^{6}-\sum_{n}Z_{n}^{A}m_{A,n}^{6}=
-m_0^2\Delta_{V\!\!A}\,.
\ee
The first two relations  are the Weinberg sum rules.
The quantity $F_{\pi}$ is the pion decay constant which is equal
in the QQM~\cite{24}:
\be
\label{f1}
F_{\pi}^2=\frac{N_fN_cM_0^2}{4\pi^2}\Lgg{}+O(1).
\ee
The residues in resonance pole contributions
in the vector and axial-vector correlators have the structure,
\be
Z_{n}^{(V,A)}=4f_{(V,A),n}^{2}m_{(V,A),n}^{2}\,,
\ee
with $f_{(V,A),n}$ being defined as electromagnetic
decay constants.

The corresponding two-point correlators for the $SU(2)$ QQM can be
calculated by variation
of external vector fields. Let us first consider the V-case. Taking
into account the external vector sources $V_{k,\mu}^i$ the Lagrangian
reads as follows:
$$
{\cal L}_{\mbox{\scriptsize aux}}^{V}=\bar{q}\left(\lefteqn{\not{\phantom{DD}}}D+
i\sum_{k=1}^2\Gamma_{V,k}^iV_{k,\mu}^i\right)q+
N_fN_{c}\Lambda^{2}\sum_{k,l=1}^{2}
\rho_{k\!,\mu}^ib_{kl}^{-1}\rho_{l\!,\mu}^i\,.
$$
After shifting the bosonic fields
$$
\rho_{k,\mu}^i\longrightarrow\rho_{k,\mu}^i-V_{k,\mu}^i\,,
$$
and integrating over fermionic degrees of freedom one comes to the
following effective action in external V-fields:
$$
S_{\mbox{\scriptsize eff}}^{V}(\rho_{k,\mu}^i,V_{k\!,\mu}^i)=-\left.N_fN_c\mbox{Tr}\ln\lefteqn
{\not{\phantom{DD}}}D\right|_{\mbox{\scriptsize reg}}+
$$
$$
+N_fN_{c}\Lambda^{2}\int d^4x\sum_{k,l=1}^{2}b_{kl}^{-1}
\left\{\rho_{k\!,\mu}^i\rho_{l\!,\mu}^i-2V_{k\!,\mu}^i\rho_{l\!,\mu}^i+
V_{k\!,\mu}^iV_{l\!,\mu}^i\right\}\,.
$$
Expanding $\left.\mbox{Tr}\ln\lefteqn{\not{\phantom{DD}}} D\right|_{\mbox{\scriptsize reg}}$
up to quadratic in fields terms, one has
$$
S_{\mbox{\scriptsize eff}}^{(2)}(\rho_{k,\mu}^i,V_{k\!,\mu}^i)=
\sum_{k\!,l=1}^{2}\left\{\frac{1}{2}\rho_{k\!,\mu}^iC_{kl}^{(\rho)\mu\nu}
\rho_{l\!,\nu}^i+N_c\Lambda^2b_{kl}^{-1}
\left[-2V_{k\!,\mu}^i\rho_{l\!,\mu}^i+
V_{k\!,\mu}^iV_{l\!,\mu}^i\right]\right\},
$$
where $C_{kl}^{(\rho)\mu\nu}$ is given by~Eq.~\gl{varder}.
Introducing the vectors
$$
\rho_{\mu}\equiv\begin{pmatrix}
  \rho_{1,\mu}^i \\
  \rho_{2,\mu}^i
\end{pmatrix}\,,\qquad V_{\mu}\equiv\begin{pmatrix}
  V_{1,\mu}^i \\
  V_{2,\mu}^i
\end{pmatrix}
$$
and taking into account~\gl{ft} (where we neglect the last term), one
integrates over $\rho_{\mu}$ with the result:
\be
\label{veff}
\frac{12\pi^2}{N_fN_c}S_{\mbox{\scriptsize eff}}(V_{\mu})=
-\frac98\left(\Lambda^4+O(\Lambda^2)\right)V_{\mu}^T\hat{H}^{\rho}_{\mu\nu}
V_{\nu}+\frac{3\Lambda^2}{4}V_{\mu}^TV_{\nu}\delta_{\mu\nu}\,,
\ee
\be
\label{cr}
\hat{H}^{\rho}_{\mu\nu}\equiv\left(\hat{A}p^2+\hat{B}^{\rho}\right)^{-1}
\hat{A}\left(\hat{B}^{\rho}\right)^{-1}(-p^2\delta_{\mu\nu}+
p_{\mu}p_{\nu})+\left(\hat{B}^{\rho}\right)^{-1}\!\!\delta_{\mu\nu}\,.
\ee
The last term in the Eq.~\gl{cr} together with that of in~\gl{veff}
form a local term which will be cancelled by the same term in the
A-case.
Substituting the identity
$$
\bar{q}\gamma_{\mu}q=\frac12\left(\bar{q}f_1(s)\gamma_{\mu}q-
\sqrt{3}\,\bar{q}f_2(s)\gamma_{\mu}q\right)\,
$$
into the vector correlator
\be
\Pi_{\mu\nu}^{\rho}(p^2)=4\int\!d^{4}x\exp(ipx)\langle
\bar{q}\gamma_{\mu}q(x)\bar{q}\gamma_{\nu}q(0)\rangle,
\ee
the latter can be rewritten through the second variational
derivatives:
\be
\label{kora2}
\Pi_{\mu\nu}^{(\rho)}(p^2)=\frac{N_fN_c}{12\pi^2}
\left[\Pi_{11}^{(\rho)}+3\Pi_{22}^{(\rho)}-
2\sqrt{3}\,\Pi_{12}^{(\rho)}\right](-\delta_{\mu\nu}p^2+p_{\mu}p_{\nu})\,,
\ee
\be
\label{pr}
\hat{\Pi}^{(\rho)}\equiv\left(\hat{A}p^2+\hat{B}^{\rho}\right)^{-1}
\hat{A}\left(\hat{B}^{\rho}\right)^{-1}.
\ee
On the other hand, this correlator is parametrized as follows
(see Eq.~\gl{cor1}):
\be
\label{kora3}
\Pi_{\mu\nu}^{(\rho)}(p^2)=\left[\frac{Z_{\rho}}{p^2+m_{\rho}^2}+
\frac{Z_{\rho'}}{p^2+m_{\rho'}^2}\right]
(-\delta_{\mu\nu}p^2+p_{\mu}p_{\nu})\,.
\ee
Comparison of \gl{kora2} and \gl{kora3} allows to obtain the
corresponding residues. 
(see Eqs.~\gl{zv1},~\gl{zv2}).
In the mean-field approximation the
vector correlator and residues can be calculated exactly which is
displayed in the Appendix.

The A-mesons must be considered together with the P-ones due to the
$P-A$ mixing. The relevant term in the effective action
$$
S_{\mbox{\scriptsize eff}}^{(2)}(\pi^i_k, a^i_{k,\mu})=\frac{1}{2}\int\frac{d^{4}p}
{(2\pi)^{4}}\sum_{k,l=1}^{2}2\pi_k^iC_{kl}^{\pi
a,\mu}a_{l,\mu}^i\,,
$$
appears by virtue of non-zero value of the corresponding second
variation of $S_{\mbox{\scriptsize eff}}(\pi^i_k, a^i_{k,\mu})$:
$$
C_{kl}^{\pi a,\mu}=-N_fN_c\int\frac{d^4q}{(2\pi)^4}
\mbox{tr}\left\{(i\gamma_{5})\frac{f_{k}\left(\left(\frac{q+p/2}{\Lambda}
\right)^2\right)}
{\!\!\not q+\frac{1}{2}\!\!\not p+iM}
(i\gamma_{\mu}\gamma_{5})\frac{f_{l}\left(\left(\frac{q+p/2}{\Lambda}
\right)^2\right)}{\!\!\not q-\frac{1}{2}\!\!\not p+iM}\right\}=\
$$
\be
=-\frac{4iN_fN_c}{(2\pi)^4}\int\frac{d^4qMf_{k}\left(\frac{q^2}{\Lambda^2}
\right)f_{l}\left(\frac{q^2}{\Lambda^2}\right)}
{\left[\left(q+\frac{1}{2}p\right)^{2}+M^{2}\right]
\left[\left(q-\frac{1}{2}p\right)^{2}+
M^{2}\right]}p_{\mu}+O\left(\frac{1}{\Lambda^2}\right).
\ee
In order to exclude the mixing terms one makes the shift of A-fields:
$$
a_{k,\mu}^i\longrightarrow a_{k,\mu}^i+D_{kl}^i\pi_l^ip_{\mu}\,,
$$
with the elements $D_{kl}^i$ being defined by the requirement of
cancellation of $P-A$ terms. This leads to changing of the kinetic
matrix~$\hat{A}^{\pi}$ due to contribution of longitudinal A-part:
\be
\label{red2}
\hat{A}^{\pi}\longrightarrow \hat{A}^{\pi}_{ren}=
\begin{pmatrix}
4\frac{m_{\rho}^2}{m_{a_1}^2}\Lgg{}+O(1) &
-\frac{\sqrt{3}}{2}+O\left(\frac{\Lgg{}}{\Lambda^2}\right)\\
-\frac{\sqrt{3}}{2}+
O\left(\frac{\Lgg{}}{\Lambda^2}\right) &
\frac32+O\left(\frac{1}{\Lambda^2}\right)
\end{pmatrix},
\ee
where the Eq.~(\ref{ma1}) and the scale~(\ref{scale}) have been
exploited. The redefinition~(\ref{red2}) does not
influence on the mass spectrum of P-mesons, but renormalizes the
pion decay constant~(\ref{f1}) is:
\be
\label{fpi2}
F_{\pi}^2 = \frac{N_fN_c M_0^2m_{\rho}^2}{4\pi^2m_{a_1}^2}
\ln\!\frac{\Lambda^2}{M_0^2}+O(1)\,.
\ee
For the model under consideration the relation~(\ref{fpi2}) fixes the
logarithm of the cutoff in terms of physical parameters.

Taking into account the point above the further calculations for the
A-case are the same as for the V-one.
As a result one finds the residues in the
meson poles. In the large-log approach one has:
\be
\label{zpi}
\tilde{Z}_{\pi}=4 F^2_{\pi} \simeq \frac{(m_{a_1}^2-m_{\rho}^2)\,\delta}
{m_{\rho}^2m_{a_1}^2m_{a_1'}^2}\,Z_1\,,\qquad
Z_1\equiv\frac{3N_fN_c\Lambda^4}{16\pi^2}\,,
\ee
\be
\label{zv1}
Z_{\rho}\simeq\frac{m_{a_1}^2}{m_{a_1}^2-m_{\rho}^2}\,4F^2_{\pi}\,,\qquad
Z_{a_1}\simeq\frac{m_{\rho}^2}{m_{a_1}^2-m_{\rho}^2}\,4F^2_{\pi}\,,
\ee
\be
\label{zv2}
Z_{\rho'}\simeq\frac{Z_1}{m_{\rho'}^2}\,,\qquad
Z_{a_1'}\simeq\frac{Z_1}{m_{a_1'}^2}\,,
\ee
where $\delta$ is given by \gl{del}.
Unlike the situation in the SP-case the residues in the VA-poles are
of the same order of magnitude:
$$
Z_{\rho}\sim Z_{a_1}\sim Z_{\rho'}\sim Z_{a_1'}=O\left(\Lambda^2\right).
$$
The statement~\gl{scale} follows from the comparison of
Eqs.~\gl{fpi2},~\gl{zpi}. The relation for~$\tilde{Z}_{\pi}$ is a
constraint on effective coupling constants of the
QQM~$\bar{\Delta}_{kl}$.

The first and the second sum rules are fulfilled
identically. The third one takes the form:
\be
\label{va3}
Z_1\left(m_{a_1'}^2-m_{\rho'}^2\right)\simeq
16\pi\alpha_s\langle\bar qq\rangle^2
\qquad \mbox{or}\qquad 3Z_1\bar{\sigma}=-\Delta_{V\!\!A}\,.
\ee
Notice that the analog of Eq.~\gl{va3} in the scalar case~\ci{24}
may be cast into the form
\be
\label{sp2}
\frac{N_fN_c}{2\pi^2}\left(m_{\sigma'}^2-m_{\pi'}^2\right)\simeq
24\pi\alpha_s\langle\bar qq\rangle^2
\qquad \mbox{or}\qquad
3Z_1\bar{\sigma}=-\frac{27}{32}\Delta_{V\!\!A}\,.
\ee
The minor discrepancy between relations \gl{va3} and \gl{sp2} is
about 16\% and can be referred to the quality of two-resonance
approximation.
The fourth sum rule looks as follows~\ci{we2}:
\be
\label{vafsr}
m_{a_1'}^2\simeq m_{\rho'}^2\simeq\frac{m_0^2}{2}
\quad\mbox{or}\quad-\frac43\bar{\Delta}_{22}\cdot3\bar{\sigma}\simeq
-\frac{m_0^2\Delta_{V\!\!A}}{2Z_1}\,.
\ee
Numerical estimations~\ci{we2} show that the last sum rule fails for
QQM with the ground and first excited sets of $V\!A$ mesons.
The structure of $Z_{\rho'}$ and $Z_{a_1'}$ shows that if $m_{a_1'}\simeq
m_{\rho'}$ then $Z_{a_1'}\simeq Z_{\rho'}$ and therefore
$f_{a_1'}\simeq f_{\rho'}$. As a consequence, these residues
approximately cancel each other in sum rules and the one-channel
results for $f_{\rho}$, $f_{a_1}$ hold~\ci{we1} (see also~\ci{we4}):
\be
f_{\rho}\simeq\frac{F_{\pi}m_{a_1}}{m_{\rho}
\sqrt{m_{a_1}^{2}-m_{\rho}^{2}}}\,,
\qquad
f_{a_1}\simeq\frac{F_{\pi}m_{\rho}}{m_{a_1}
\sqrt{m_{a_1}^{2}-m_{\rho}^{2}}}\,.
\ee
After evaluating we get $f_{\rho}\approx 0.15$ and $f_{a_1}\approx
0.06$ to be compared with the experimental values
$f_{\rho}=0.20\pm 0.01$,
$f_{a_1}=0.10\pm 0.02$~\cite{we1}.

It should be mentioned that introducing VA-meson fields influences
on some CSR constraints for SP-case obtained in~\cite{24}.
Due to the redefinition~\gl{red2}
the first SP sum rule is not fulfilled now identically but up to
terms of order $O(1/\Lambda^2)$ because of the scaling~\gl{scale}. The
second SP sum rule
is not changed in the large-$\Lambda$ approximation.
The relation $Z_{\sigma}\simeq Z_{\pi}$ is not now valid.
Instead one has
$$
Z_{\sigma}\simeq\frac{m_{\rho}^2}{m_{a_1}^2}Z_{\pi}\,.
$$
The relevant physical discussions and fits for the SPVA QQM can be traced
in~\ci{we1,we2}. From the estimations~\ci{we2}, in particular,
follows that $m_{a_1'}-m_{\rho'}\approx 60$ MeV which proves a
fast restoration of chiral symmetry.

\section{Summary}

\begin{enumerate}
\item
We have shown that $SU(2)$ and $U(3)$ versions of Quasilocal Quark Model
with chirally invariant four-fermion vector and axial-vector
vertices including derivatives in fields can serve to describe the physics
of vector and axial-vector meson resonances at intermediate energies.
The corresponding
mass spectrum for the ground and first excited VA boson states was derived
in the mean field and large-log approximations. The qualitative features of
the mass spectrum obtained turned out to be the same as in the
scalar-pseudoscalar case~\cite{24}: the excited states are logarithmically
heavier than ground ones and a fast restoration of chiral symmetry over
the scale 1~GeV is predicted. Comparison with the SP, $SU(2)$ QQM permitted
to obtain two relations between boson masses independent of model
parameters:
$$
m_{a_1}^{2} - m_{\rho}^{2} \simeq \frac32m_{\sigma}^2\,,
$$
$$
m_{a_1'}^{2} - m_{\rho'}^{2}\simeq\frac32(m_{\sigma'}^2-m_{\pi'}^2).
$$
The latter relation predict the mass of the first radial excitation of
axial-vector meson in the energy range $m_{a_1}=1500\div1550$ MeV.
\item
The VA, $U(3)$ generalization of QQM allowed to derive much more relations
between masses of meson states which do not depend on any model parameters
(see Eqs.~\gl{u28}-\gl{u31}).
The agreement with experimental data is within 10\% for the V-case and
15\% for the A-one~\ci{we2}.
\item
From the expansion of QCD two-point colour-singlet current
correlators for VA-fields in inverse powers of large momentum and
the comparison with OPE one obtains the set of sum rules for the
differences of VA-correlators, which show a rapid decrease at large
momenta. Therefore, the chiral symmetry is restored at high
energies and one can perform the QCD matching by means of few
lowest lying VA-resonances that gives a set of constraints on
parameters for VA, $SU(2)$ QQM. In particular, the residues for the ground
and excited VA states turned out to be of the same order of magnitude unlike
the situation in the SP case. The inclusion of excited states did not
change the electromagnetic decay constants
of the ground VA states compared with the one-channel results:
$f_{\rho}\approx 0.15$, $f_{a_1}\approx 0.06$.
\item
The main results of our work have been obtained by means of method
of effective action and on the basis of the idea of chiral
symmetry restoration at high energies and on OPE of the two-point
correlators for vector and axial-vector quark densities. All
calculations have been performed in the large-$N_c$ and
log-approximations.
\item
We conclude that the VA Quasilocal Quark Model reflects
phenomenology of low and intermediate energy meson physics and
the matching to
non-perturbative QCD based on chiral symmetry restoration at high
energies improves the predictability of the models.

\end{enumerate}

\section*{Acknowledgements}

We are grateful to A. A. Andrianov for the numerous fruitful
discussions and attention to our work.
This work is supported by Grant RFBR
01-02-17152, INTAS Call 2000 Grant (Project 587), and The Program
"Universities of Russia: Fundamental Investigations" (Grant UR.02.01.001).

\section*{Appendix}

In this Appendix we calculate in the mean field approach the two-point
vector correlator for the two-channel QQM. After introducing
external vector fields $V_{k,\mu}^i$ and integrating over $\rho_{k,\mu}^i$
one has for the quadratic in fields part of effective action:
$$
\frac{12\pi^2}{N_fN_c}S_{\mbox{\scriptsize eff}}^{(2)}(V_{\mu})=
-\frac98\Lambda^4V_{\mu}^T\hat{K}\left(\hat{A}p^2+\hat{B}^{\rho}\right)^{-1}
\hat{A}\left(\hat{B}^{\rho}\right)^{-1}\hat{K}V_{\nu}(-p^2\delta_{\mu\nu}+
p_{\mu}p_{\nu})+
$$
\be
\label{s2}
+\frac{3}{4}\Lambda^2V_{\mu}^T\left(\hat{K}-\frac32\Lambda^2\hat{K}
\left(\hat{B}^{\rho}\right)^{-1}\hat{K}\right)V_{\nu}\delta_{\mu\nu}\,,
\ee
where
$$
\hat{K}=\begin{pmatrix}
  1-\frac{4\bar{\Delta}_{11}}{3\Lambda^2} &
  -\frac{4\bar{\Delta}_{12}}{3\Lambda^2} \\
  -\frac{4\bar{\Delta}_{12}}{3\Lambda^2} &
  1-\frac{4\bar{\Delta}_{22}}{3\Lambda^2}
\end{pmatrix}.
$$

In order to obtain the correlator of local currents~\gl{cor1} one
needs to project the expression~\gl{s2} on the local current:
$$
\Pi^{V}_{\mu\nu}(p^{2})=\mbox{\rm tr}\left\{\hat{P}\frac{\delta^2S_{\mbox{\scriptsize eff}}^{(2)}}
{\delta V_{\mu}\delta V_{\nu}}\right\},\qquad
\hat P = \begin{pmatrix}
  1 & -\sqrt{3} \\
  -\sqrt{3} & 3
\end{pmatrix},
$$
where $\hat P$ is the projection operator. It is a formal way to
obtain the relation~\gl{kora2}.
The final result is:
\ba
\label{fin}
\Pi^{V}_{\mu\nu}(p^{2})&=&\frac{Z_1\left(ap^2+b+c\right)
(-p^2\delta_{\mu\nu}+p_{\mu}p_{\nu})}{24\left(\Lgg{}-2
\right)\det\hat{\bar{\Delta}}\left(p^2+m_{\rho}^2\right)\left(p^2+
m_{\rho'}^2\right)}+\nl
&+&\frac34\Lambda^2\left(\frac{3d\Lambda^2}{4\det\hat{\bar{\Delta}}}-4\right)
\delta_{\mu\nu}\,.
\ea
In the Eq.~\gl{fin} we introduced the following notations:
$$
a\equiv -6\left(\Lgg{}-2\right)\left[d+8\varepsilon\det\hat{\bar{\Delta}}+
\varepsilon^2\det\hat{\bar{\Delta}}(\bar{\Delta}_{11}+3\bar{\Delta}_{22}-
2\sqrt{3}\bar{\Delta}_{12})\right],
$$
$$
b\equiv 8\left(\Lgg{}-2\right)\left(\bar{\Delta}_{22}+
\sqrt{3}\bar{\Delta}_{12}+\varepsilon\det\hat{\bar{\Delta}}\right)^2,
$$
$$
c\equiv \left(d+4\varepsilon\det\hat{\bar{\Delta}}\right)^2,\qquad
\varepsilon\equiv -\frac{4}{3\Lambda^2}\,,
$$
and $d$, $Z_1$ are given by~\gl{d},~\gl{zpi}.
The residues in poles of the transverse part of
correlator~\gl{fin} are:
\be
\label{ezr1}
Z_{\rho}=\frac{Z_1\left(-am_{\rho}^2+b+c\right)}{24\left(\Lgg{}-2
\right)\det\hat{\bar{\Delta}}\left(m_{\rho'}^2-m_{\rho}^2\right)}
\ee
\be
\label{ezr2}
Z_{\rho'}=\frac{Z_1\left(-am_{\rho'}^2+b+c\right)}{24\left(\Lgg{}-2
\right)\det\hat{\bar{\Delta}}\left(m_{\rho}^2-m_{\rho'}^2\right)}
\ee
One can check that the expressions \gl{ezr1}, \gl{ezr2} in the
large-log approximation turn into corresponding
quantities~\gl{zv1},~\gl{zv2}.

The exact calculation of axial-vector correlator is much more difficult
because of $P-A$ mixing.

\renewcommand{\refname}{References}


\begin{thebibliography}{99}
\bibitem{11} A. A.~Andrianov, V. A.~Andrianov, --- Int. J. Mod.
Phys. {\bf A8} (1993), 1981;
Theor. Math. Phys. {\bf 94} (1993), 3;
in: Proc. School-Sem. "Hadrons and nuclei from
QCD'', Tsuruga/Vladivostok/Sapporo 1993, ed. by K. Fujii et al.
World Scientific, Singapore,
1994, p. 341; hep-ph/9309297.
\bibitem{1} Y.~Nambu, G.~Jona-Lasinio, --- Phys. Rev. {\bf 122}
(1961), 345.
\bibitem{2} M. K.~Volkov,--- Ann. Phys. (N. Y.) {\bf 157} (1984), 282.
\bibitem{3} U.-G.~Meissner, --- Phys. Rept. {\bf 161} (1988), 213.
\bibitem{4} H.~Vogl, W.~Weise, --- Progr. Part. Nucl. Phys. {\bf
27} (1991), 195.
\bibitem{5} S.~Klevansky, --- Rev. Mod. Phys.  {\bf 64} (1992), 649.
\bibitem{6} A. A.~Andrianov, V. A.~Andrianov, --- Z. Phys.  {\bf
C55} (1992), 435;
Theor. Math. Phys. {\bf 93} (1992), 1126;
J. Math. Sci. (N. Y.) {\bf 77}  (1995), 3021 (Zap. Nauch. Sem.
POMI {\bf 199} (1992), 3).
\bibitem{7} J.~Bijnens, C.~Bruno, E.~de Rafael, --- Nucl. Phys.
{\bf B390} (1993), 501.
\bibitem{8} T.~Hatsuda, T.~Kunihiro, --- Phys. Rept.
{\bf 247} (1994), 221.
\bibitem{9} D.~Ebert, H.~Reinhardt, M. K.~Volkov, ---
Progr. Part. Nucl. Phys. {\bf 33} (1994), 1.
\bibitem{10} J.~Bijnens, --- Phys. Rept. {\bf 265} (1996), 369.
\bibitem{pdg} K. Hagiwara et al., --- Phys. Rev. {\bf D66} (2002), 010001.
\bibitem{12}  A. A.~Andrianov, V. A.~Andrianov, ---
Nucl. Phys. Proc. Suppl. {\bf 39BC} (1995), 257.
\bibitem{13} E.~Pallante, R.~Petronzio, --- Z.~Phys. {\bf C65}
(1995), 487.
\bibitem{14} A. A.~Andrianov, V. A.~Andrianov, V. L.~Yudichev, ---
Theor. Math. Phys. {\bf 108} (1996), 1069.
\bibitem{volw} M. K.~Volkov, C.~Weiss, --- Phys. Rev.
{\bf D56} (1997), 221.
\bibitem{volkeb}
M. K. Volkov, D. Ebert, M. Nagy, --- Int. J. Mod. Phys.
{\bf A13} (1998), 5443.
\bibitem{15} D.~Espriu, E.~de Rafael, J.~Taron, --- Nucl. Phys.
{\bf B345} (1990), 22; erratum ibid. {\bf B355} (1991), 278.
\bibitem{16} G.~t'Hooft, --- Nucl. Phys. {\bf B72} (1974), 461; E.~Witten,
Nucl. Phys. {\bf B160} (1979), 57.
\bibitem{18} M. A.~Shifman,
A.I.~Vainstein, V. I.~Zakharov, --- Nucl. Phys. {\bf B147} (1979), 385, 448.
\bibitem{zap} A. A.~Andrianov, V. A.~Andrianov, --- Zap. Nauch.
Sem. POMI {\bf 245} (1996), 5; hep-ph/9705364.
\bibitem{20} M.~Knecht, E.~de Rafael, --- Phys. Lett. {\bf B424}
(1998), 335.
\bibitem{21} S.~Peris, M.~Perrottet, E.~de Rafael, --- JHEP {\bf
05} (1998), 011.
\bibitem{iof} B. L.~Ioffe, K. N.~Zyablyuk, --- hep-ph/0010089.
\bibitem{we1} A. A.~Andrianov, V. A.~Andrianov, S. S.~Afonin, ---
Proc. of the 15th Int. Workshop on High Energy Physics and Quantum
Field Theory (Tver, 2000), ed. by M. N.~Dubinin and V. I.~Savrin, Moscow,
2001, p. 233; hep-ph/0101245.
\bibitem{24} A. A.~Andrianov, V. A.~Andrianov, ---
Proc. of the Int. Workshop on Hadron
Physics (Coimbra, 1999), ed. by A. H. Blin et al., N. Y., AIP, 2000, p. 328;
hep-ph/9911383.
\bibitem{we3} A. A.~Andrianov, V. A.~Andrianov, S. S.~Afonin, ---
Proc. of the 16th Int. Workshop on High Energy Physics and Quantum
Field Theory (Moscow, 2001), ed. by M. N.~Dubinin and V. I.~Savrin, Moscow,
2002, p. 254.
\bibitem{we2} A. A.~Andrianov, V. A.~Andrianov, S. S.~Afonin, ---
Proc. of the 12th Int. Sem. on High Energy
Physics QUARKS'2002 (Novgorod, 2002), to be published; hep-ph/0209260.
\bibitem{we4} V. A. Andrianov, S. S. Afonin, --- Yad. Fiz. {\bf 65}
(2002), 1; hep-ph/0109026.
\end{thebibliography}
\end{document}